# FLARES OF SAGITTARIUS A* AT MILLIMETER WAVELENGTHS

**Atsushi Miyazaki** [(1) †], Takahiro Tsutsumi [(2)], Makoto Miyoshi [(2)], Masato Tsuboi [(3)], and Zhi-Qiang Shen [(1)]

[(1)] *Shanghai Astronomical Observatory, Chinese Academy of Sciences, 80 Nandan Road, Shanghai 200030, P.R. China;* [†] *e-mail: amiya@shao.ac.cn,*

[(2)] *National Astronomical Observatory of Japan, 2-21-1 Osawa, Mitaka, Tokyo 181-8588, Japan,*

[(3)] *Nobeyama Radio Observatory, NAOJ, Nobeyama, Minamimaki, Minamisaku, Nagano 384-1305, Japan.*

## ABSTRACT

We have performed monitoring observations of the flux density toward the Galactic center compact radio source, Sagittarius A* (Sgr A*), which is a supermassive black hole, from 1996 to 2005 using the Nobeyama Millimeter Array of the Nobeyama Radio Observatory, Japan. These monitoring observations of Sgr A* were carried out in the 3- and 2-mm (100 and 140 GHz) bands, and we have detected several flares of Sgr A*. We found intraday variation of Sgr A* in the 2000 March flare. The twofold increase timescale is estimated to be about 1.5 hr at 140 GHz. This intraday variability suggests that the physical size of the flare-emitting region is compact on a scale at or below about 12 AU (≈150 $R$s; Schwarzschild radius). On the other hand, clear evidence of long-term periodic variability was not found from a periodicity analysis of our current millimeter data set.

## INTRODUCTION

Sagittarius A* (Sgr A*) is a unique compact radio source located at the dynamical center of the Galaxy, and is associated with a supermassive black hole (SMBH) of mass ~4×10$^6$ $M_\Theta$ (e.g., [9]). After its discovery, Sgr A* was observed at many wavelengths, and temporal flux variations were reported. However, its emission mechanisms are not yet fully understood. Because this source is embedded in thick thermal material, it is practically difficult to observe its detailed structure. Time variability observation is a powerful tool to probe the structure and emission mechanisms, if the variability is intrinsic.

At centimeter wavelengths, the time variability of Sgr A* has been studied over the past two decades (e.g., [5] [20] [10]). Very Large Array (VLA) monitoring at short cm-wavelengths suggests a periodicity of about 110 days [20] [21]. However, at cm-wavelengths, Sgr A* is dominated by effects of refractive interstellar scintillation, and it is difficult to measure all of its intrinsic properties. These effects are usually weaker at higher frequencies, so we can better ascertain the intrinsic nature of Sgr A* at millimeter wavelengths. Earlier, in 1990, significant flux variations of Sgr A* at 89 GHz were reported by [19]. Subsequently, we detected a flare in Mar 1998 from the Sgr A* flux monitoring observations at mm-wavelengths, using the Nobeyama Millimeter Array (NMA) at the Nobeyama Radio Observatory (NRO) since 1996 [13] [17]. We also detected a few flares after 1998 [14] [18]. Similar flares of Sgr A* were also detected at 1-mm wavelength using the Sub-Millimeter Array [22]. Recently, X-ray and infrared flares of Sgr A* were also detected, and these indicate very short timescales and violent intensity increases [1] [16] [7] [8].

At mm-wavelengths, Sgr A* is a relatively weak (~1 Jy) compact component embedded in the extended and strong HII region of Sgr A West. It is necessary to observe with higher angular resolution (≤ a few arcsec) to discriminate the compact component from the extended components. Therefore, we have conducted the flux monitoring experiments of Sgr A* at short millimeter wavelengths using the NMA at the NRO.

## OBSERVATIONS AND CALIBRATIONS

We have performed flux monitoring observations of Sgr A* in the 100 and 140 GHz bands (λ= 3 and 2-mm) using the NMA, a six element 10-m dish interferometer, at the NRO from 1996 to 2005. The observations were carried out using multiple array configurations of the NMA over a period of a few months to half a year for each observable season. Each epoch consists of a set of sequential observations over about 2 days. The epochs were separated by several days to 2 weeks. For each observing day, the observing time was 2–4 hr within the maximum observable time of about 4 hr.

We used double sideband (DSB) SIS receivers in the 3- and 2-mm bands as the front-ends. The data in 1996 were obtained using the FX correlator with 320 MHz bandwidth at 102 and 146 GHz. All the data after 1997 were obtained using the Ultra Wide Band Correlator (UWBC) with 1 GHz bandwidth. With this UWBC system, simultaneous observations of both the lower and upper sidebands, separated by 12 GHz, are allowed. Thus, these observed frequencies were 90 and 102 GHz for the 3-mm band, and 134 and 146 GHz for the 2-mm band. The instrumental gain and phase were calibrated by alternating observations of Sgr A* and NRAO530 at about 20 min intervals. We also used an additional phase calibrator, 1830-210, a known QSO, after 2000. The flux densities of the calibrators are determined from Uranus or Neptune, which were used as the primary flux calibrators. The absolute uncertainties of the flux scaling are about 15% and 20% at the 100 and 140 GHz bands, respectively. Because the observations for Sgr A* in the 2-mm band requires the best weather conditions and phase stability, these observations were made less frequently.

Most of the observations, including the detections of the flares of Sgr A*, were performed by the array configuration with intermediate baselines, C-configuration, of the NMA. The projected baselines in this array range over ~7–55 k$\lambda$ in the 100 GHz band and ~10–77 k$\lambda$ in the 140 GHz band. We made spatially filtered maps from the visibilities with projected baselines exceeding 25 kilo-wavelengths (($U^2+V^2)^{1/2} \geq 25$ k$\lambda$) to suppress the contamination from the extended components surrounding Sgr A* (17 k$\lambda$ for the data taken with the compact array configuration, D-configuration, in the 100 GHz band). Typical synthesized beam sizes (HPBW) for the C-configuration were about 3"×6" and 2"×4" at 100 and 140 GHz, respectively. From the flux densities of the calibrators measured on the maps, the fractions of the decorrelation due to atmospheric phase fluctuation are 10–20% at 100 GHz and 20–40% at 140 GHz. We corrected the measured flux densities of Sgr A* for the decorrelation, and averaged two measured flux densities of Sgr A* which were individually calibrated by each of the two phase calibrators. A more detailed description of the observations and the data calibrations have also been presented in our other papers [14] [15] [18].

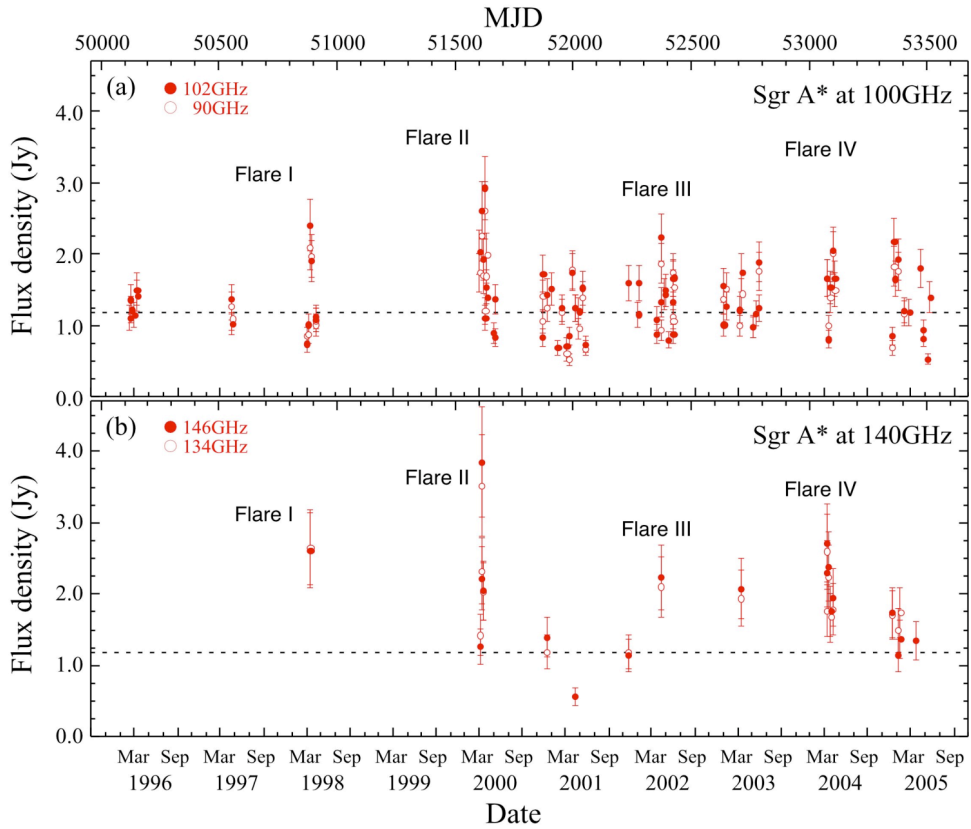

Fig.1, Light curves of Sgr A* at (a) 100 and (b) 140 GHz bands from 1996 to 2005. There are at least four active phases in Mar 1998, Mar 2000, Apr 2002, and Mar 2004 (Flare I, Flare II, Flare III, and Flare IV, respectively). The dashed lines indicate the mean quiescent flux density.

**RESULTS AND DISCUSSION**

Fig. 1 shows the light curves of Sgr A* at 100 and 140 GHz bands constructed from all the data from 1996 to 2005. The total number of observations is about 90 days. These light curves show that Sgr A* has quiescent and active phases [14]. The mean flux densities in the quiescent phase are 1.1±0.2 Jy and 1.2±0.2 Jy at 90 and 102 GHz, respectively. There are at least four flares in March 1998, March 2000, April 2002, and March 2004 (Flare I, Flare II, Flare III, and Flare IV, respectively, indicated in Fig. 1) during the whole observing period [18]. The peak flux density in these flares increases by 100%–200% at 100 GHz, and by 200%–400% at 140 GHz ($\Delta S/S$), respectively, while the averaged quiescent flux density was about 1 Jy. It appears that the flux density variability increases with frequency [14] [17].

Fig. 2a shows the light curves of Sgr A* at 100 and 140 GHz in March 2000 [14] [15], and indicates particularly strong variability in the active phase of Flare II. The most prominent flare was observed on 7 March 2000 in the 140 GHz band, and the peak flux density at 146 GHz was 3.9±0.8 Jy. The flux density on the subsequent day, 8 March, decreased to 2.2±0.4 Jy at 146 GHz. The half flux decay timescale of the flare at 146 GHz was at most 24 hours. We divided the data set in the 140 GHz band observed on 7 and 8 Mar 2000 into about 5-15 minute bins, and measured the flux density of Sgr A* in each bin in order to search for shorter timescale variability. Fig. 2b shows the light curves of Sgr A* in the 140 GHz bund over the two days [15]. On 7 March, the flux density around the peaks changed rapidly. The flux density of Sgr A* at 146 GHz increased from 3.5 to 4.7 Jy between 21:45 to 22:15 UT on 7 March. The peak flux densities were 4.2±0.8 Jy at 134 GHz, and 4.7±0.9 Jy at 146 GHz. The typical relative uncertainties within one observation session of the 100 and 140 GHz bands were estimated to be a few % and 6%, respectively. The upper panels in Fig. 2b show the light curves of the calibrators (NRAO530, 1830-210). The scatters of the measured flux densities are much smaller than the variation in the flare of Sgr A*. Thus, the observed 30% increase in 30 minutes must be real. The timescale that the flux density increased by 100% (two-fold increase timescale) is estimated to be about 1.5 hours assuming that the increase has a constant gradient [15]. Our result is consistent with the intraday variation detected at 3-mm using the OVRO mm-interferometer in May 2002 [11]. Intraday variation of Sgr A* has also been reported at cm-wavelengths [4]; however, its amplitude is much smaller than that in our mm-observations. The intraday flare at mm-wavelengths has a similar increase timescale as those known in the X-ray and infrared flares, but has smaller amplitude. The short increase timescale, 1.5 hr, suggests that the physical size of the flare-emitting region in the accretion disk is compact on a scale at or below about 12 AU (≈150 $R$s; the Schwarzschild radius $R$s = $2GM/c^2$, assuming a black hole mass of $4\times10^6\,M_\odot$) [15]. There is asymmetry in the light curve as the half decay timescale is much longer than the flare increase timescale. A light curve with rapid increase and slow decay, as obtained in our observations, is similar to that often observed in outburst phenomena with ejections, e.g., flares on the Sun, the 1972 outburst of Cyg X-3 [12], etc.

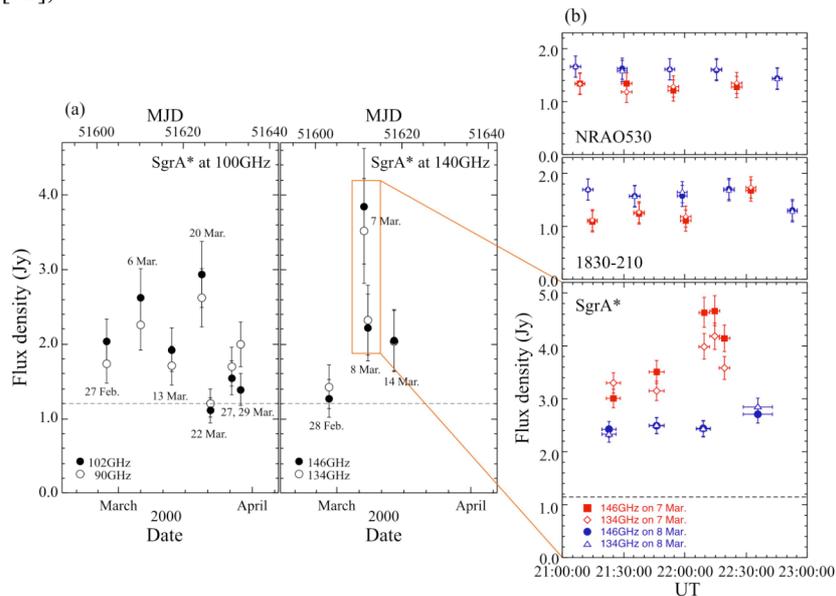

Fig.2, (a) Light curve of Sgr A* at 100 (*left panel*) and 140 GHz (*right panel*) in March 2000 [14] [15]. The flux density was violently changing. There is a prominent peak in the 140 GHz band on 7 March. (b) Light curve of the Sgr A* flare (*bottom panel*) and the calibrators (*upper panels*), NRAO530 and 1830-210, at 140 GHz on 7 (*red*) and 8 (*blue*) March 2000 [15]. The flux density of Sgr A* at 146 GHz (*red filled squares in bottom panel*) increased from 3.5 to 4.7 Jy, from 21:45 to 22:15 UT on 7 March. In both (a) and (b), the horizontal dashed lines indicate the mean flux density in the quiescent phase.

A decaying phase of Sgr A* was observed in Mar 2004. In Flare IV, the flux densities on 6 Mar 2004 at 134 and 146 GHz were 2.6±0.5 Jy and 2.7±0.5 Jy, respectively. The flux densities in the 2-mm band decreased on a few observing days after 6 Mar, so Sgr A* clearly appeared to be in a decaying phase. During this phase, we had an observation at 140 GHz on 31 Mar 2004, which was just before the X-ray flare detected with *XMM-Newton* [3], and the flux density was at a slightly higher level of ~2 Jy, compared to the mean quiescent flux density. The flux density on the subsequent day,

1 Apr, was also at a slightly higher level of ~2 Jy at 100 GHz. The relations among X-ray, infrared, and radio flares should be important for probing the emission mechanisms of Sgr A*. Apparent correlations between radio and X-ray flares [23], and between infrared and X-ray flares [6], were also reported. However, there are only a few of our observations which overlap the reported X-ray or infrared flares. Some X-ray flares of Sgr A* detected in the *Chandra* observations from 22 to 30 May 2002 were reported [2]. Although we also carried out some NMA observations during this period, we detected no significant mm-flare in unfavorable weather conditions.

Reference [20] reported the presence of a 106 day periodicity in the centimeter flux variability of Sgr A* from the analysis of data observed with the VLA over 20 years. We showed that the millimeter light curve exhibits distinct high and low activity states, by folding our NMA data with this period of 106 days [14]. These results may indicate a quasi-periodic flux variability of Sgr A* at mm-wavelengths. Then we carried out a periodicity analysis using the Lomb-Scargle periodogram method for our NMA data at 3-mm [18]. However, no significant peak in the normalized power spectra is found for the entire 3-mm data set. For the subset of these data after 2000, the power spectra has a peak at $0.0101 d^{-1}$ corresponding to a ~99 day period, with a significance level of 0.37 (false alarm probability). Thus, from the analysis, this periodicity is not significant for our data [18]. Our current sampling is probably insufficient to detect a periodicity, if we consider possible fluctuations in the period, as indicated in recent analysis of VLA cm-data [21].

## SUMMARY

We performed Sgr A* flux monitoring in the 100 and 140 GHz bunds using the NMA at the NRO from 1996 to 2005.
1) We have detected several flares of Sgr A* at 100 and 140 GHz during the whole observing period. The peak flux density of the flares increases by 100%–200% at 100 GHz, and by 200%–400% at 140 GHz ($\Delta S/S$), respectively, while the averaged quiescent flux density was ~1 Jy. The variability of flux density increases with frequency.
2) We found intraday variation of Sgr A* in the 2000 March flare. The twofold increase timescale is estimated to be about 1.5 hr at 140 GHz. This intraday variability suggests that the physical size of the flare-emitting region is compact on a scale at or below about 12 AU ($\approx$150 $R$s; Schwarzschild radius).
3) Long term periodic variability is not clear from the periodicity analysis for our current mm-data. Our current data sampling is probably insufficient, if we consider the possibility of fluctuations in the period.

## ACKNOWLEDGEMENTS
We thank the NMA staff (NRO) for observational support, and Dr. A.B. Fletcher (SHAO) for help with proofreading.

## REFERENCES
[1] Baganoff, F.K., Bautz, M.W., Brandt, W.N., et al. 2001, *Nature*, 413, 45
[2] Baganoff, F.K., et al. 2002, AAS 201st Meeting, #31.08, *BAAS*, 34, 1153
[3] Bèlanger, G., Goldwurm, A., Renaud, M., et al. 2005, *ApJ*, submitted (astro-ph/0508128)
[4] Bower, G.C., Falcke, H., Sault, R.J., & Backer, D.C. 2002, *ApJ*, 571, 843
[5] Brown, R.L., & Lo, K.Y. 1982, *ApJ*, 253, 108
[6] Eckart, A., et al. 2004, *A&A*, 427, 1
[7] Genzel, R., Schödel, R., Ott, T., et al. 2003, *Nature*, 425, 934
[8] Ghez, A.M., Wright, S.A., Matthews, K., et al. 2004, *ApJL*, 601, L159
[9] Ghez, A.M., Salim, S., Hornstein, S.D, et al. 2005, *ApJ*, 620, 744
[10] Herrnstein, R.M., Zhao, J.-H., Bower, G.C., & Goss, W.M. 2004, *AJ*, 127, 3399
[11] Mauerhan, J.C., Morris, M., Walter, F., & Baganoff, F.K. 2005, *ApJL*, 623, L25
[12] Martí, J., Paredes, J.M., & Estalella, R. 1992, *A&A*, 258, 309
[13] Miyazaki, A.,Tsutsumi, T., & Tsuboi, M. 1999, *Advances in Space Research*, 23, 5/6, 977
[14] Miyazaki, A.,Tsutsumi, T., & Tsuboi, M. 2003, *Astron. Nachr.*, 324, S1, 363
[15] Miyazaki, A.,Tsutsumi, T., & Tsuboi, M. 2004, ApJL, 611, L97
[16] Porquet, D., Predehl, P., Aschenbach, B. et al. 2003, *A&A*, 407, L17
[17] Tsuboi, M., et al. 1999, in *The Central Parsecs of the Galaxy*, ASP Conf. Ser. 186 (San Francisco: ASP), 105
[18] Tsutsumi, T., Miyazaki, A., & Tsuboi, M. 2005, *AJ*, submitted
[19] Wright, M.C.H., & Backer, D.C. 1993, *ApJ*, 417, 560
[20] Zhao, J.-H., Bower, G.C., & Goss, W.M. 2001, *ApJL*, 547, L29
[21] Zhao, J.-H. 2003a, *Astron. Nachr.*, 324, S1, 355
[22] Zhao, J.-H., Young, K.H., Herrnstein, R.M. et al. 2003b, *ApJL*, 586, L29
[23] Zhao, J.-H., Herrnstein, R.M., Bower, G.C., Goss, W.M., & Liu, S.M. 2004, *ApJL*, 603, L85